\documentclass[english, prb, showpacs, reprint]{revtex4-1}
\usepackage[latin9]{inputenc}
\usepackage{color}
\usepackage{babel}
\usepackage{verbatim}
\usepackage{bm}
\usepackage{amsmath}
\usepackage{amssymb, mathtools}
\usepackage{graphicx}
\usepackage[unicode=true,pdfusetitle,
 bookmarks=true,bookmarksnumbered=false,bookmarksopen=false,
 breaklinks=false,pdfborder={0 0 0},pdfborderstyle={},backref=false,colorlinks=true]
 {hyperref}
\hypersetup{
 citecolor=blue}

\makeatletter
\usepackage{epstopdf}
\usepackage{amsfonts}
\usepackage{mathrsfs}\usepackage{bm}\usepackage{appendix}
\usepackage{txfonts}
\usepackage{float}
\usepackage{epstopdf}
\usepackage{ulem}

\makeatother

\begin{document}

\preprint{APS/123-QED}

\title{Electronic structure, magnetism and high-temperature superconductivity in the multi-layer octagraphene and octagraphite}

%\thanks{Thankes}%

\author{Jun Li$^{1}$}
\author{Shangjian Jin$^{1}$}
\author{Fan Yang$^{2}$}
\author{Dao-Xin Yao$^{1}$}%
\email{yaodaox@mail.sysu.edu.cn}
\affiliation{%
	$^{1}$State Key Laboratory of Optoelectronic Materials and Technologies, School of Physics, Sun Yat-Sen
	University, Guangzhou 510275, Peoples Republic of China\\
	$^2$School of Physics, Beijing Institute of Technology, Beijing 100081, China
}%

\date{\today}% It is always \today, today,
%  but any date may be explicitly specified

\begin{abstract}
We systematically investigate the electronic structure, magnetism and high-temperature superconductivity (SC) in the multi-layer octagraphene and octagraphite (bulk octagraphene). A tight binding model is used to fit the electronic structures of single-layer, multi-layer octagraphenes and octagraphite. We find that the multi-layer octagraphene and octagraphite follow a simple A-A stacking structure from the energy analysis. The van der Waals interaction induces $t_{\perp}\approx0.25$ eV and the hopping integrals within each layers changes little when the layer number $n$ increases. There is a well Fermi-surface nesting with nesting vector $\mathbf{Q}=(\pi,\pi)$ for the single-layer octagraphene at half-filling, which can induce a 2D N\'{e}el antiferromagnetic order. With increasing the layer number $n\rightarrow\infty$, the Fermi-surface nesting transforms to 3D with nesting vector $\mathbf{Q}=(\pi,\pi,\pi)$ and shows the system has a 3D N\'{e}el antiferromagnetic order. Upon doping, the multi-layer octagraphene and octagraphite can enter a high-temperature $s^{\pm}$ SC driven by spin fluctuation. We evaluate the superconducting transition temperature $T_c$ by using the random-phase approximation (RPA), which yields a high $T_c$ even if the layer number $n\geq$ 3. Our study shows that the multi-layer octagraphene and octagraphite are promising candidates for realizing the high-temperature SC.
\end{abstract}

\maketitle

\section{\label{sec:1}INTRODUCTION}
The two-dimensional (2D) superconductors have drawn tremendous interests for their rich physical properties and potential applications. So far, the SC has been reported in many 2D materials, such as FeSe-SrTiO$_3$~\cite{Wang_2012}, monolayer NbSe$_2$~\cite{Lu_2015}, MoS$_2$~\cite{Xi_2016}, CuO$_2$~\cite{Zhu_2016}, Bi$_2$Sr$_2$CaCu$_2$O$_{8+\delta}$~\cite{Yu_2019},  $etc$. As the first single-layer 2D material, graphene shows an interesting proximity-induced superconductivity when it contacts with SC materials~\cite{Heersche_2007}. Besides, few-layer graphene with doping may exhibit a considerable superconducting transition temperature $T_c$~\cite{Xue_2012,Li_2013,Ludbrook_2015,Tiwari_2017,Huder_2018}, which is higher than the reported $T_c$ in bulk compounds of the same composition~\cite{Calandra_2005}. Recently, the ``high-temperature SC" with a $T_c\sim 1.7$ K has been revealed in the magic-angle twisted bi-layer graphene~\cite{Cao_2018}. These progresses inform us that combinations and interactions between layers may bring important influence to the properties of 2D materials.

Theoretically, the SC of graphene-based 2D materials has been widely studied via the Eliashberg theory under the framework of electron-phonon coupling mechanism (BCS)~\cite{Calandra_2012,Jelena_2014,Kaloni_2013,Mazin_2010,Si_2013,Wang_2018}. By doping and applying a biaxial stress, the highest $T_c$ of graphene-based materials has been proposed to reach 30 K~\cite{Si_2013}. In addition to graphene, variable forms of graphyne have been predicted and some were synthesized~\cite{Malko_2012}. It is only predicted that $\alpha$-graphyne would exhibit a SC with $T_c\sim 12$ K by hole-doping and biaxial tensile strain~\cite{Toktam_2016}. The hexagon symmetry of graphene or graphyne is unfavorable to form the Fermi surface nesting with high density of states, which is important to form the high-temperature superconductivity.

Another 2D carbon-based material is the octagraphene~\cite{Liu_2012,Sheng_2012}. Astonishingly, the 2D square-octagon lattice structure of the single-layer octagraphene leads to a high density of states near the well-nested Fermi-surface (FS), which may induce an antiferromagnetic spin-density-wave (SDW) order. The BCS mechanism based on electron-phonon interaction is not enough to describe the pairing and the SC mainly originates from spin fluctuation. Our recent research on a repulsive Hubbard model on a square-octagon lattice with nearest-neighbor and next-nearest-neighbor hopping terms, which can serve as a rough representation of the single-layer octagraphene, shows that the system can host the high-temperature SC with $s^{\pm}$-wave pairing symmetry~\cite{Kang_2019}. Unlike the complex forms of other 2D superconductors, the simple structure of octagraphene may be an ideal platform for studying the origin of high-temperature SC. In real materials, multi-layer octagraphene and octagraphite may be more common. We here attend to study the electronic structures, magnetism and high-temperature superconductivity in the multi-layer octagraphene and octagraphite.

Meanwhile, the synthesizations of octagraphene, multi-layer octagraphene and octagraphite are in progress. While a novel synthesization route of single-layer octagraphene has been proposed theoretically~\cite{Gu_2019}, an one-dimensional carbon nanoribbons with partial four and eight-membered rings has been realized experimentally~\cite{Liu_2017}. As octagraphene shows a low cohesive energy~\cite{Sheng_2012}, it has an opportunity to build the strongest carbon atomic sheet after graphene.

In this paper, we get a better tight binding (TB) model model to study the band structure of single-layer octagraphene. In comparison with our previous work~\cite{Kang_2019}, the present Hamiltonian adopts hopping integrals fitted from the density-functional theory (DFT) calculations and are thus more realistic. Unlike the complex stacking of the graphene, our DFT calculation suggests that multi-layer octagraphenes build more likely an A-A stacking. There is a well Fermi-surface nesting with nesting vector $\mathbf{Q}=(\pi,\pi)$ for the single-layer octagraphene at half-filling, which can induce a 2D N\'{e}el antiferromagnetic order. With increasing the layer number $n\rightarrow\infty$, the Fermi-surface nesting transforms to 3D with nesting vector $\mathbf{Q}=(\pi,\pi,\pi)$ and shows the system has a 3D N\'{e}el antiferromagnetic order. Upon doping, the multi-layer octagraphene and octagraphite can enter a high-temperature $s^{\pm}$ SC driven by spin fluctuation. We calculate the $T_c$ of single-layer octagraphene, multi-layer octagraphene, and octagraphite, and find that the interlayer interaction would not affect the superconducting state much. With increasing the $n$, $T_c$ converges to $\sim170$ K, which is still high.

The rest of the paper is organized as follows. In sec.~\ref{sec:method} we provide our model and the details of our methods. In Sec.~\ref{sec:single}, we introduce the calculation to single-layer octagraphene and compare with our previous work. In Sec.~\ref{sec:multi}, we study the property of multi-layer octagraphenes.  Sec.~\ref{sec:bulk} provides the results for octagraphite, which is different from the multi-layer octagraphenes. The exhibited $T_c$ with increasing the layer number $n$ is given in our estimation. Finally, in Sec.~\ref{sec:con} we provide the conclusions.

\section{\label{sec:method} Model and Approach}

\subsection{\label{sec:Model} The Model}

We use the projector augmented wave (PAW) method implemented in Vienna ab initio simulation package (VASP) to perform the density functional theory (DFT) calculations~\cite{vasp1,vasp2,vasp3,vasp4}. The generalized gradient approximation (GGA) and the Perdew Burke-Ernzerhof (PBE) function are used to treat the electron exchange correlation potential~\cite{Perdew_1996}. The vacuum is set as 15 $\AA$ to avoid the external interaction. Grimme's DFT-D3 is chosen to correct the van der Waals interaction~\cite{Grimme_2010}. An extremely high cutoff energy (1500 eV) and 16$\times$16$\times$1 k-point mesh with Monkhorst-Pack scheme are used in the self-consistent calculation.

\begin{figure}[t]
	\centering
	\includegraphics[width=0.5\textwidth]{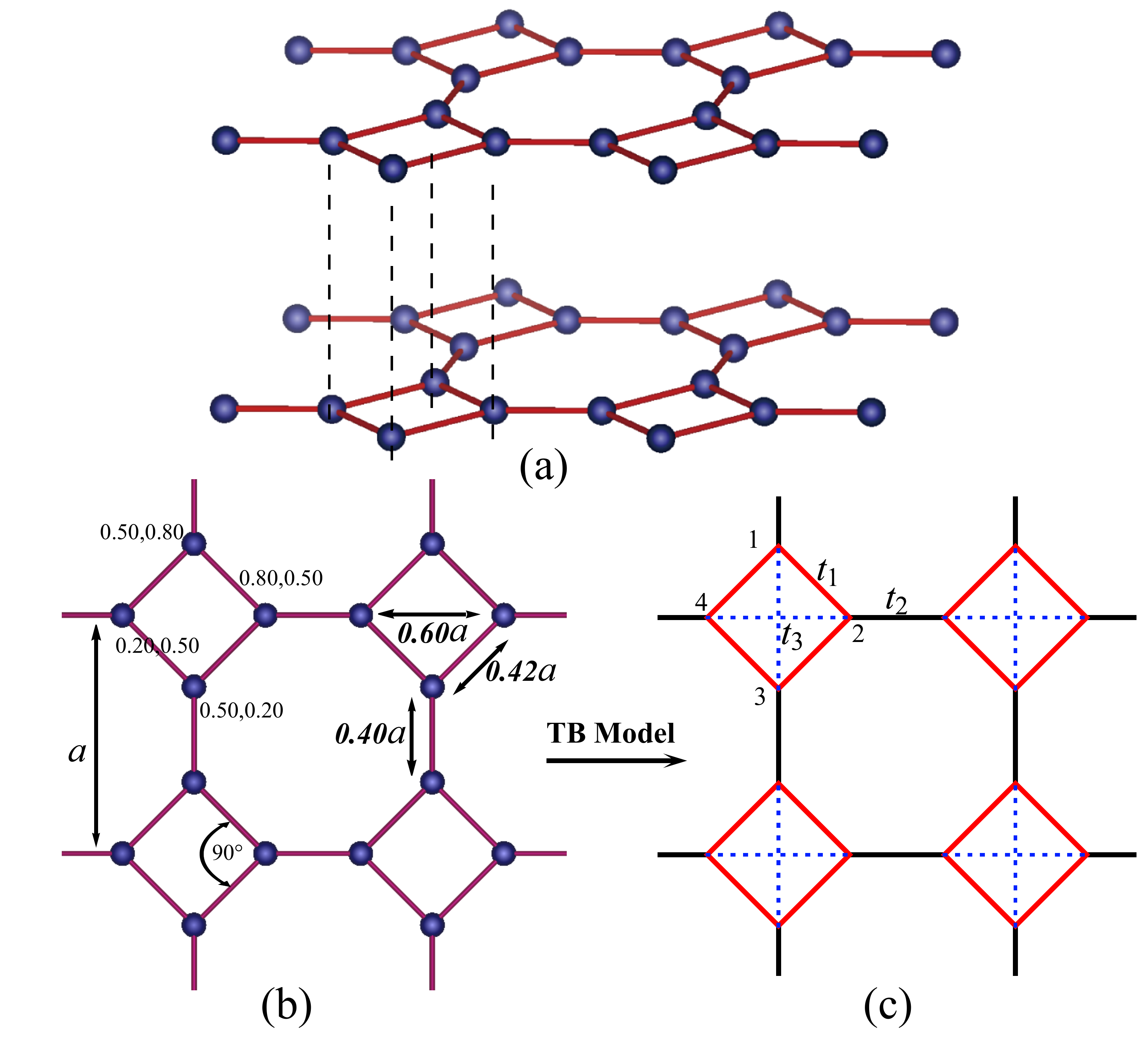}
	\caption{\label{fig:model} (a) The predicted structure of octagraphene from DFT calculation. The relative positions between the layers form the A-A stacking. (b) Structure of single-layer octagraphene. The relative positions of four carbon atoms in a unit cell are independent of the deformation. (c) 2D single-orbital tight binding (TB) model. $t_1$, $t_2$ and $t_3$ correspond to the intra-square, inter-square and diagonal hopping energies, respectively.}
\end{figure}

To quantitatively analyze the band structures from DFT calculations, we build a tight binding (TB) model to describe the single-layer octagraphene, multi-layer octagraphene and octagraphite. The Hamiltonian can be expressed as
\begin{eqnarray}
\label{eq:tb}
H_{TB}=-\sum_{i, j, \sigma} t_{ij} c_{i \sigma}^{\dagger} c_{j \sigma}-\sum_{<i, j>}t_{\perp}c_{i}^{\dagger} c_{j}+H.c.,
\end{eqnarray}
where $c_{i \sigma}^{\dagger}$ ($c_{i \sigma}$) is the electron creation (annihilation) operator for a given site $i$ with spin $\sigma$. $t_{ij}$ is the hopping energies defined in Fig.~\ref{fig:model}(c) and $t_{\perp}$ represents the Van der Waals interlayer interaction between neighbor layers. Note that the matrix form of Eq. (\ref{eq:tb}) is different for the single-layer octagraphene, multi-layer octagraphene and octagraphite.

Similarly as graphene, there are strong Coulomb repulsions between the $2p_z$ electrons in the octagraphene materials. Here we use an effective Hubbard model to describe the effects
\begin{equation}
\label{eq:U}
H_{Hubbard}=H_{\mathrm{TB}}+U \sum_{i} \hat{n}_{i \uparrow} \hat{n}_{i \downarrow}.
\end{equation}
Here the $U$-term represents the on-site repulsive Hubbard interaction between the $2p_z$ electrons within the same site.
\subsection{\label{sec:approach} The RPA approach}

We use the procedure of RPA outlined in our prior work~\cite{Kang_2019,Liu_2013} to solve Eq. (\ref{eq:U}). With generally neglecting the frequency dependence, we define free susceptibility for $U = 0$
\begin{equation}
\label{eq:chi}
\chi_{s, t}^{(0) p, q}\left(\mathbf{q}\right)=\frac{1}{N} \sum_{\mathbf{k}, \alpha, \beta} \xi_{t}^{\alpha}(\mathbf{k}) \xi_{s}^{\alpha, *}(\mathbf{k}) \xi_{q}^{\beta}(\mathbf{k'}) \xi_{p}^{\beta, *}(\mathbf{k'}) \frac{n_{F}\left(\varepsilon_{\mathbf{k'}}^{\beta}\right)-n_{F}\left(\varepsilon_{\mathbf{k}}^{\alpha}\right)}{\varepsilon_{\mathbf{k}}^{\alpha}-\varepsilon_{\mathbf{k'}}^{\beta}}.
\end{equation}
where $\alpha,\beta=1,2,3,4$ are band indices, $\mathbf{q}=\mathbf{k'}-\mathbf{k}$ is the nesting vector between $\mathbf{k'}$ and $\mathbf{k}$, $\varepsilon_{\mathbf{k}}^{\alpha}$ and $\xi_{\xi}^{\alpha}(\mathbf{k})$ are the $\alpha$th eigenvalue and eigenvector of matrix form of Eq. (\ref{eq:tb}) respectively and $n_F$ is the Fermi-Dirac distribution function.

In the RPA level, the spin (charge) susceptibility for the Hubbard-model is
\begin{equation}
\label{eq:chis}
\chi^{(c(s))}(\mathbf{q})=\left[I+(-)\chi^{(0)}(\mathbf{q})\widetilde{U}
\right]^{-1} \chi^{(0)}(\mathbf{q})
\end{equation}
where $\chi^{(c(s))}(\mathbf{q})$, $\chi^{(0)}(\mathbf{q})$ and $\widetilde{U}$ are $16 \times 16$ matrices with $\widetilde{U}_{s t}^{p q}=U \delta_{s=t=p=q}$.

A Cooper pair with momentum $\mathbf{k'}$and orbital $(t, s)$ could be scattered to $\mathbf{k}$, $(p, q)$ by charge or spin fluctuations. In the RPA level, to project the effective interaction into the two bands which cross the Fermi surface, we obtain the following low energy effective Hamiltonian for the Cooper pairs near the Fermi surface,
\begin{equation}
\label{eq:v}
V_{e f f}=\frac{1}{N} \sum_{\alpha \beta, \mathbf{k k}^{\prime}} V^{\alpha \beta}\left(\mathbf{k}, \mathbf{k}^{\prime}\right) c_{\alpha}^{\dagger}(\mathbf{k}) c_{\alpha}^{\dagger}(-\mathbf{k}) c_{\beta}\left(-\mathbf{k}^{\prime}\right) c_{\beta}\left(\mathbf{k}^{\prime}\right),
\end{equation}
where $\alpha,\beta = 1, 2$ and $V^{\alpha \beta}$ is
\begin{equation}
V^{\alpha \beta}\left(\mathbf{k}, \mathbf{k}^{\prime}\right)=\operatorname{Re} \sum_{p q s t, \mathbf{k} \mathbf{k}^{\prime}} \Gamma_{s t}^{p q}\left(\mathbf{k}, \mathbf{k}^{\prime}, 0\right) \xi_{p}^{\alpha, *}(\mathbf{k}) \xi_{q}^{\alpha, *}(-\mathbf{k}) \xi_{s}^{\beta}\left(-\mathbf{k}^{\prime}\right) \xi_{t}^{\beta}\left(\mathbf{k}^{\prime}\right).
\end{equation}

In the singlet channel, the effective vertex $\Gamma_{s t}^{p q}\left(k, k^{\prime}\right)$ is given as follow,
\begin{equation}
\label{eq:lam}
\begin{aligned}
	\Gamma_{s t}^{p q}\left(k, k^{\prime}\right)=\widetilde{U}_{q s}^{p t}&+\frac{1}{4}\left\{\widetilde{U}\left[3 \chi^{(s)}\left(k-k^{\prime}\right)-\chi^{(c)}\left(k-k^{\prime}\right)\right]\widetilde{U}\right\}_{q s}^{p t}\\&+
	\frac{1}{4}\left\{\widetilde{U}\left[3 \chi^{(s)}\left(k+k^{\prime}\right)-\chi^{(c)}\left(k+k^{\prime}\right)\right]\widetilde{U}\right\}_{q t}^{p s},
\end{aligned}
\end{equation}
while in the triplet channel, it is
\begin{equation}
\label{eq:Tc}
\begin{aligned}
\Gamma_{s t}^{p q}\left(k, k^{\prime}\right)=&-\frac{1}{4}\left\{\widetilde{U}\left[\chi^{(s)}\left(k-k^{\prime}\right)+\chi^{(c)}\left(k-k^{\prime}\right)\right]\widetilde{U}\right\}_{q s}^{p t}\\&+
\frac{1}{4}\left\{\widetilde{U}\left[\chi^{(s)}\left(k+k^{\prime}\right)+\chi^{(c)}\left(k+k^{\prime}\right)\right]\widetilde{U}\right\}_{q t}^{p s}.
\end{aligned}
\end{equation}

We can construct the following linear integral gap equation to determine the $T_c$ and the leading pairing symmetry of the system from low energy effective Hamiltonian Eq. (\ref{eq:v})
\begin{equation}
-\frac{1}{(2 \pi)^{2}} \sum_{\beta} \oint_{F S} d k_{\|}^{\prime} \frac{V^{\alpha \beta}\left(\mathbf{k}, \mathbf{k}^{\prime}\right)}{v_{F}^{\beta}\left(\mathbf{k}^{\prime}\right)} \Delta_{\beta}\left(\mathbf{k}^{\prime}\right)=\lambda \Delta_{\alpha}(\mathbf{k}).
\end{equation}
Here, the integration and summation are along variable Fermi surface patches labeled by $\alpha$ or $\beta$. The $v_{F}^{\beta}$ is Fermi velocity at $k'$ on the $\beta$th Fermi surface patch, and $\mathbf{k'},\mathbf{k}$ represent the component along that patch. In the eigenvalue problem, the normalized eigenvector  $\Delta_{\alpha}(\mathbf{k})$ represents the relative value of the gap function on the $\alpha$th Fermi surface patch. The largest pairing eigenvalue  $\lambda$ is used to estimate $T_c$ by the following equation,

\begin{equation}
\lambda^{-1}=\ln \left(1.13 \frac{\hbar \omega_{D}}{k_{B} T_{c}}\right),
\end{equation}
here we all choose the typical energy scale of spin fluctuation $\hbar\omega_{D}$ = 0.3 eV in our calculation, see reference~\cite{Liu_2013}.

\section{\label{sec:single} single-layer octagraphene }
\begin{figure}[t]
	\centering
	\includegraphics[width=0.5\textwidth]{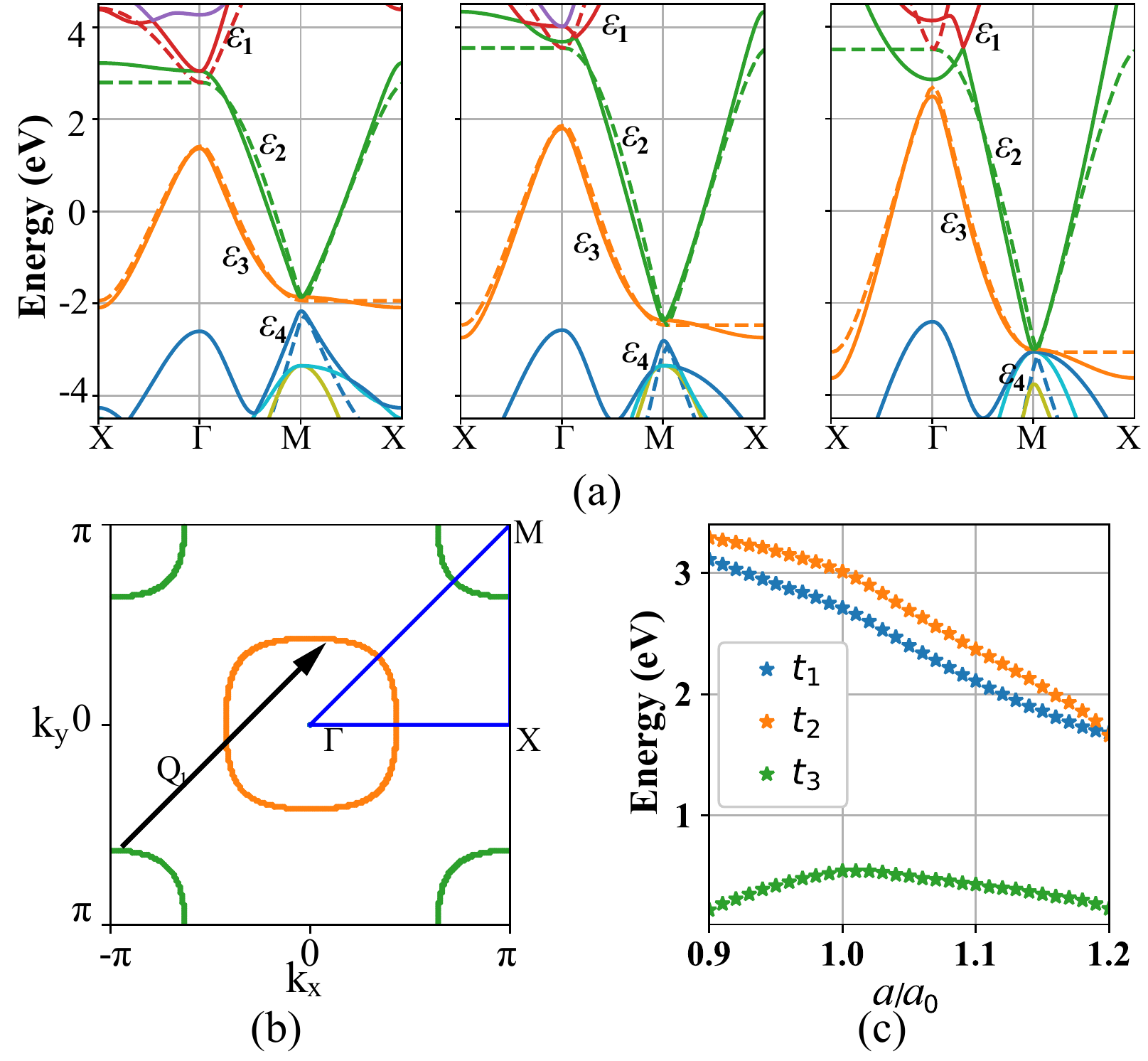}
	\caption{\label{fig:single}Single-layer octagraphene. (a) Band structures of different lattice constant $a$: $a/a_0$ = 1.1, 1.0 and 0.9 [$a_0=3.44$ $\AA$]. DFT calculated results, solid lines; fitting results obtained by TB model, dashed lines. For $a/a_0=0.9$, the bands show a quadruple degeneracy at the $M$ point with $E = -3.01$ eV. (b) Fermi surface from TB model, independent of the relative lattice constant $a/a_0$. The Fermi surface is well nested by the vector $\mathbf{Q}_1 = (\pi,\pi)$. (c) Variable fitting parameters $t_1$, $t_2$ and $t_3$ of TB model with lattice constant $a$. $t_2/t_1=1.1$ is almost constant independent of $a$.}
\end{figure}

In our DFT calculation of single-layer octagraphene, the fit of Brich-Murnaghan EOS gives the more accurate lattice constant $a_0$ = 3.44 $\AA$. We note that the relative positions of carbon atoms are almost independent of the lattice constant $a$.
%%The calculated square-octagon lattice is shown in Fig. \ref{fig:model}(b). The positions of four carbon atoms in the unit cell are at (0.50$a$, 0.20$a$), (0.20$a$, 0.50$a$), (0.50$a$, 0.80$a$) and (0.80$a$, 0.50$a$) respectively, consistent with $C_{4v}$ symmetry. Similar as graphene, each carbon atom of octagraphene forms a $sp^2$ hybridization with its nearest three carbon atoms, two of which form the square edges (0.42$a$) while the last one forms octagonal edge (0.40$a$). One $s$ orbital and two $p$ orbitals form the $\sigma$ bonds.
The rotational symmetry of $\sigma$ bonds of octagraphene are lower than graphene, and hence the octagraphene is less stable than graphene. The rest $p$ orbital electrons form the $\pi$ bonds similar as the graphene.

In Fig.~\ref{fig:single}(a), we show our DFT calculated band structures with variable lattice constant $a$. There are two bands $\epsilon_2$ and $\epsilon_3$ near the Fermi level. %%For strained larger lattice constant $a$, the $\epsilon_2$ and $\epsilon_3$ become gentler and separate from other bands. However, the Stress reduces the lattice constant $a$ and leads to precipitous band structures and make $\epsilon_2$ and $\epsilon_3$ overlap with other bands.
For $a/a_0=0.9$, the bands are quadruplely degenerate at the $M$ point with $E$ = -3.01 eV. This coincidence is different from the Dirac point. The structure is not a bi-conical structure with linear dispersion, but a parabolic dispersion. It means low-energy excitations are no-longer massless. %%This shape of dispersion is also found in the multi-layer octagraphenes with $n\ge 2$.
%%While it may be fruitful to investigate the reason, we leave this topic for future studies.

At the Fermi level, the band structures contain a hole pocket around the $\Gamma$ point and an electron pocket around the $M$ point, see Fig.~\ref{fig:single}(b). This is similar to the undoped Fe-pnictides materials~\cite{Hirschfeld_2011}. The two pockets connected by the nesting vector $\mathbf{Q_1}=(\pi,\pi)$ form the well Fermi-surface nesting, which is independent of deformations within the single-layer.

After a general procedure of Fourier transformation, the Hamiltonian Eq. (\ref{eq:tb}) of single-layer reads as
\begin{equation}
\label{eq:band}
\widetilde{H}_1=-\left[\begin{array}{cccc}{0} & {t_1} & {t_2 e^{i k_{y}}+t_3} & {t_1} \\ {t_1} & {0} & {t_1} & {t_2 e^{i k_{x}}+t_3} \\ {t_2 e^{-i k_{y}}+t_3} & {t_1} & {0} & {t_1} \\ {t_1} & {t_2 e^{-i k_{x}}+t_3} & {t_1} & {0}\end{array}\right].
\end{equation}

We obtain four bands $\epsilon_1$, $\epsilon_2$, $\epsilon_3$ and $\epsilon_4$ by diagonalizing Eq. (\ref{eq:band}). Since the $\epsilon_1$ and $\epsilon_4$ are away from the Fermi level, we only use the $\epsilon_2$ and $\epsilon_3$  to get better fittings. By fitting the bands $\epsilon_2$ and $\epsilon_3$ of the path from $\Gamma$ to $M$ points, we get $t_1$ = 2.678 $\pm$ 0.033 eV, $t_2$ = 2.981 $\pm$ 0.027 eV and $t_3$ = 0.548 $\pm$ 0.024 eV with $a/a_0$ = 1.0. In comparison, $t\approx$ 2.7 eV of nearest-neighbor hopping energy and $t^{\prime}\approx$ 0.1 eV of next nearest-neighbor hopping energy is reported in graphene~\cite{Castro_2009}. Note that the existence of this small $t_3$ is necessary to split the $\epsilon_3$ and $\epsilon_4$ at $M$ point, and make $\epsilon_2$ coincides with $\epsilon_3$ here.

%%The sizes and positions of the hole pocket around the $\Gamma$ point and the electron pocket around the $M$ point
$\mathbf{Q_1}$ remains almost unchanged with different deformations, see Fig.~\ref{fig:single}(b). This is due to that the diagonalization result of Eq. (\ref{eq:band}) is mathematically independent of deformation $a/a_0$. This phenomenon is also examinated by our DFT calculation, supporting the credibility of our TB model. Such an unchanged Fermi-surface nesting may stabilize the SC phase of the octagraphene.

Figure~\ref{fig:single}(c) shows the variable fitting parameters $t_1$, $t_2$ and $t_3$ of TB model with lattice constant $a$. As the distances between carbon atoms enlarge, the values of $t_1$, $t_2$ and $t_3$ decrease. This leads to the flatter band structures in Fig.\ref{fig:single}(a). However, $t_2/t_1$ remains almost $1.1$ when $a$ changes from $0.90a_0$ to $1.20a_0$. The relative interaction $t_2/t_1$ is independent of $a$. We may conclude that the hopping energies between carbon atoms are nearly inversely proportional to distances based on our calculations.

We then use a Hubbard model in Eq. (\ref{eq:U}) to study the influence of spin fluctuation on SC. Although the interaction parameter $U$ would be more than 10 eV for the graphene-based materials, the accurate value of $U$ is still under discussion~\cite{Castro_2009}. Due to the weak-coupling character of RPA, there is a limitation for the value of $U$, i. e. $U_c$. Here, we set $U$ = 5.4 eV (2$t_1$) and have the electron doping density $x$ as 10$\%$ according to our estimation of the limits of RPA. The details of RPA limitation $U_c$ will be elaborated in Sec.~\ref{sec:bulk}. The diagonalizing  eigen-susceptibilities $\chi(\mathbf{q})$ of Eq. (\ref{eq:chi}) peaks at the vector $\mathbf{Q_1}=(\pi,\pi)$, also verified by our DFT result. The related eigenvector of susceptibilities $\epsilon(\mathbf{Q}_1)$ = $(1/2,-1/2,1/2,-1/2)$ means that the N\'{e}el pattern is formed, see Fig.~\ref{fig:bulk}(d).

We then get $\lambda$ = 0.321 for $a/a_0$ = 1.0 and $T_c$ $\sim$ 190 K for the single-layer octagraphene. For comparison, it has been reported recently that the calculated $T_c$ is $20.8$ K within the framework of electron-phonon coupling ~\cite{Gu_2019}. Our calculated $T_c$ is much higher due to the spin fluctuation, not the electron-phonon interaction. In the previous study, our variational Monte Carlo gives the superconducting gap amplitude $\Delta$ $\sim$ 50 meV and the similar $T_c$ at $\sim$ 180 K with the $s^\pm$-wave pairing~\cite{Kang_2019}. The consistence between the two methods shows great chance to search for high $T_c$ superconductor.

We also note that with the decreasing of $a$, $T_c$ decreases in a limited scale. This may be explained by the weakness of interactions. However, $T_c$ would remain a high value ($>$ 100 K) when $a/a_0$ from 0.9 to 1.2. Thus single-layer octagraphene would be a good superconductor with limited mechanical deformation.

\section{\label{sec:multi}Multi-layer octagraphene}

\begin{figure}[t]
	\centering
	\includegraphics[width=0.5\textwidth]{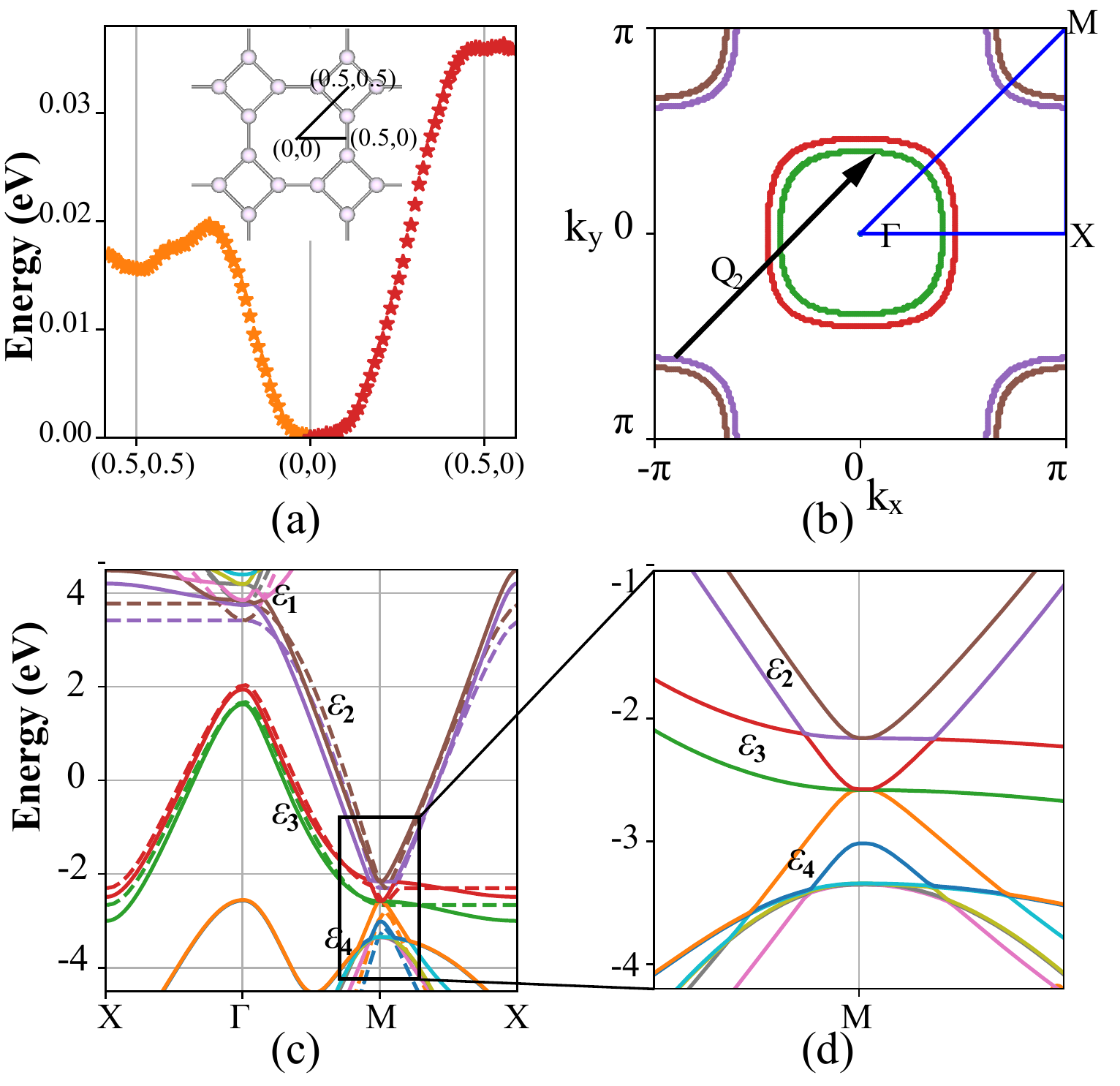}
	\caption{\label{fig:multi} (a) The differences between cohesive energy per atom of the bi-layer octagraphene with relative shifts. The relative shifts between the two layers are chosen along the $(100)$ and $(110)$ in real space. A-A stacking (0,0) is the most stable in our calculation. (b) Fermi surface of bi-layer octagraphene. The nesting vectors $\mathbf{Q}_2$ = $(\pi,\pi)$, $(\pi+\delta,\pi+\delta)$ and $(\pi-\delta,\pi-\delta)$ mean the deviation of perfect Fermi surface nesting. (c) Band structures of the bi-layer octagraphene with $a_0$ = 3.45 $\AA$. The solid lines represent the results by DFT calculation. The dashed lines are fitting results of TB model. (d) The detailed bands near the $M$ point. Three branches from $\epsilon_2$, $\epsilon_3$ and $\epsilon_4$ coincide and form a triple degeneracy at the $M$ point.}
\end{figure}
In real materials, multi-layer octagraphene may be more common. We here apply a DFT$+$RPA method to study the properties of multi-layer octagraphenes. We firstly verify the stacking modes of bi-layer octagraphene. Due to the  $C_{4v}$ symmetry of single-layer, there may be three mostly possible stacking modes between two octagraphene layers: A-A stacking, A-B stacking and A-C stacking, which are defined as (0, 0), (0.5, 0.5) and (0, 0.5) relative shifts between the two layers, respectively. The differences between cohesive energy per atom along $(100)$ and $(110)$ directions are shown in Fig.~\ref{fig:multi}(a). In our calculations, the A-A (0,0) stacking is the most stable. Otherwise, from A-A (0,0) stacking to A-B (0.5, 0.5) stacking, the energy differences are smaller compared with graphene. The distance between the neighboring layers of multi-layer octagraphene is 3.72 $\AA$ , which is larger than the value of graphene ($3.4 \AA$). This indicates a weaker inter-layer coupling, making the material more slippery than graphite~\cite{Liu_2014}.

Since the A-A stacking bi-layer is the most stable stacking mode, we only consider the A-A stacking structure. The bi-layer Hamiltonian near the Fermi surface in matrix form reads as
\begin{equation}
\label{eq:band2}
\widetilde{H}_2=\left[\begin{array}{cc}{\widetilde{H}_1} & {t_{\perp}\widetilde{I}_{4\times4}} \\ {t_{\perp}\widetilde{I}_{4\times4}} & \widetilde{H}_1 \end{array}\right],
\end{equation}
where $\widetilde{H}_1$ is Eq. (\ref{eq:band}),  $\widetilde{I}_{4\times4}$ is a $4\times4$ identity matrix.

The fitting parameters of bi-layer octagraphene are $t_1$ = 2.685 $\pm$ 0.021 eV, $t_2$ = 3.001 $\pm$ 0.016 eV, $t_3$ = 0.558 $\pm$ 0.016 eV and $t_{\perp}$ = 0.184 $\pm$ 0.011 eV. $t_1$, $t_2$ and $t_3$ have little deference from single-layer octagraphene. This can be understood by the small interlayer interaction $t_{\perp}$, smaller than that of graphene ($t_{\perp}$ $\approx$ 0.4 eV)~\cite{Castro_2009}. However, each band of single-layer splits into two bands due to the doubled unit cell. As a result, there are two nesting hole pockets around the $\Gamma$ point and two nesting electron pockets around the $M$ point, seen Fig.~\ref{fig:multi}(b).

Interestingly, three branches from $\epsilon_2$, $\epsilon_3$ and $\epsilon_4$ coincide and form a triple degeneracy at the $M$ point, see Figs.~\ref{fig:multi}(c) and (d). This triple degeneracy, which naturally exists in the bi-layer octagraphene, does not need any external deformation. From our TB model, the diagonalizing of Eq. (\ref{eq:band2}) gives the exactly same result at the $M$ point when $t_1+t_{\perp}=t_2+t_3$ is satisfied. While matching of single-layer $\epsilon_2$ and $\epsilon_3$ at the $M$ point is determined by the $C_{4v}$ symmetry, the matching with $\epsilon_4$ is just a coincidence.

The usage of RPA for bi-layer octagraphene gives $\lambda$ = 0.324 for $U$ = 5.4 eV, doping $x=$ $10\%$, which has a little difference from single-layer octagraphene. We obtain $T_c$ $\sim$ 180 K, which is a bit lower than that in single-layer octagraphene. We suppose that this may be caused by the interlayer interaction and the cell expansion. Although $t_{\perp}$ is very small compared with the intralayer interactions, the well Fermi-surface nesting of one layer is deviated by the interlayer interaction, see Fig.~\ref{fig:multi}(b). There are two hole and two electron pockets with the nesting vectors $\mathbf{Q}_2$ = $(\pi,\pi)$, $(\pi+\delta,\pi+\delta)$ and $(\pi-\delta,\pi-\delta)$. The bluring of perfect Fermi surface nesting suppresses the superconductivity and reduces the $T_c$.

Then we tend to study the tendency of SC with increasing the layer number $n$.  The A-A stacking multi-layer octagraphenes show more 2D-like behavior.  As the $n$ increases, the two energy bands $\epsilon_2$ and $\epsilon_3$ split into more branches due to the expansion of unit cell. We can still use the same form of Eq. (\ref{eq:band2}), which can be written as:
\begin{equation}
\label{eq:band3}
\widetilde{H}_n=\left[\begin{array}{cccc}{\widetilde{H}_1} & {t_{\perp}\widetilde{I}_{4\times4}} &{}&{0}\\
{t_{\perp}\widetilde{I}_{4\times4}} & {\widetilde{H}_1}&{t_{\perp}\widetilde{I}_{4\times4}}&{}\\
{}&{\ddots}&{\ddots}&{\ddots}\\{0}&{}&{t_{\perp}\widetilde{I}_{4\times4}} & {\widetilde{H}_1}
\end{array}\right].
\end{equation}

We fit the DFT calculated data of $\epsilon_2$ and $\epsilon_3$ of the path from $\Gamma$ to $M$ points to Eq. (\ref{eq:band3}). The fitting parameters and $\lambda$ of tri- to six- layer are reported in Table~\ref{tab:1}. We find that the fitting parameters are very close to those of bi-layer octagraphene, whose relative difference are all less than one percent.

With increasing the layer number $n$, we find that the pairing symmetry is kept unchanged as s$^{\pm}$, and the $T_c$ does not change much. According to our estimation, we get $T_c$ $\sim$ 170 K  for tri- to five- layer and about $T_c$ $\sim$ 160 K for six-layer when $U$ = 5.4 eV, doping $x=$ $10\%$. Thus we suggest superconductivity of octagraphene is related to the 2D characteristics of materials.

\section{\label{sec:bulk}Octagraphite}

\begin{figure}[htbp]
	\centering
	\includegraphics[width=0.5\textwidth]{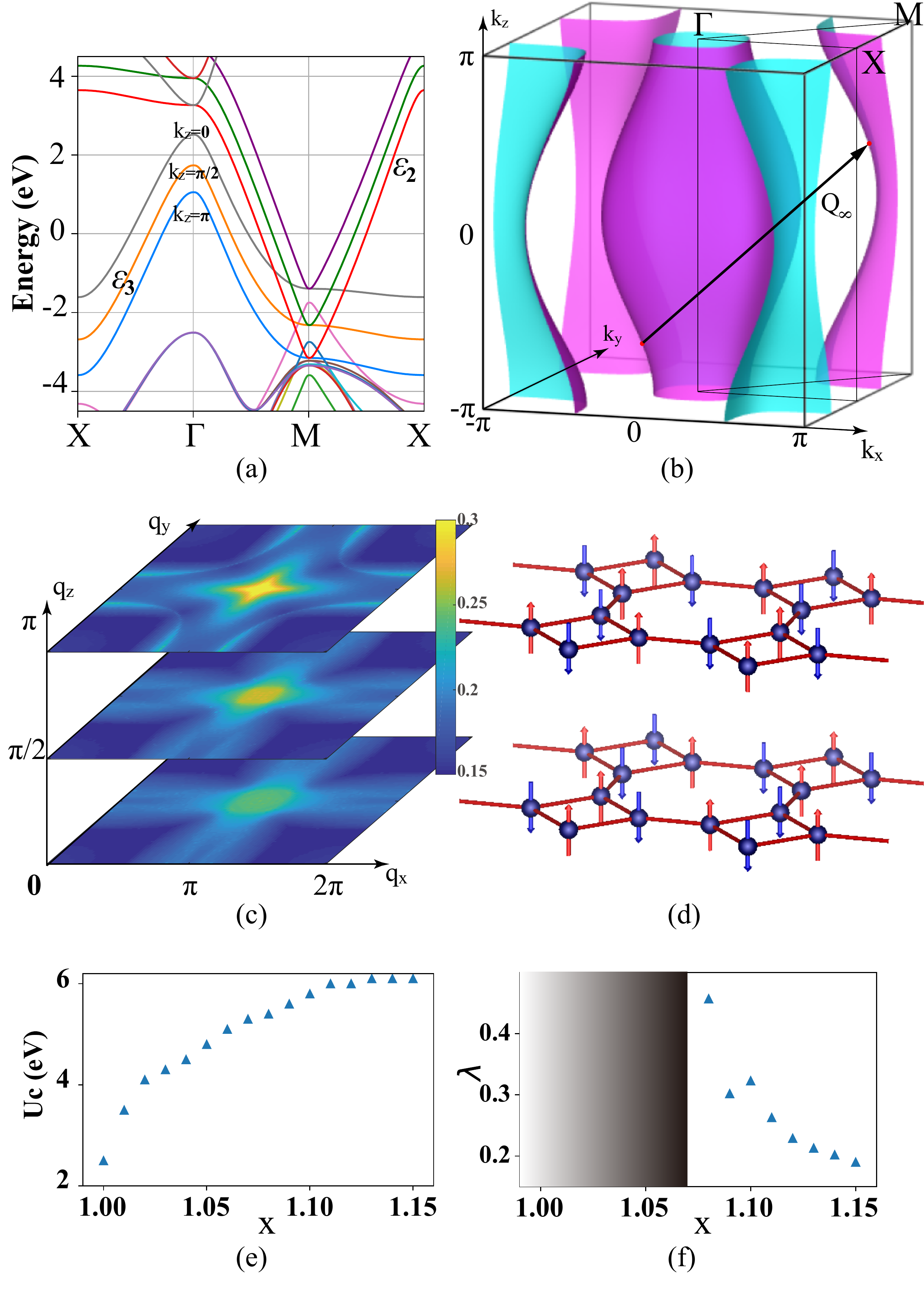}
	\caption{\label{fig:bulk} Octagraphite. (a) Band structures with $k_z =$ 0, $\pi/2$, $\pi$. (b) Fermi surface obtained by VESTA~\cite{Momma_2011}, the nesting vector is almost $\mathbf{Q}_\infty=(\pi,\pi,\pi)$. (c) The eigen-susceptibilities $\chi(\mathbf{q})$ with $\mathbf{q}_z$ = 0, $\pi/2$, $\pi$. $\chi(\mathbf{q})$ peaks at almost $\mathbf{Q}_\infty=(\pi,\pi,\pi)$. (d) Predicted antiferromagnetic N\'{e}el pattern with half filling. (e) RPA calculated $U_c$ as a function of the electron doping density $x$. (f) Doping density $x$ dependent of largest pairing eigenvalues $\lambda$ with $U$ = 5.4 eV. Based on (e) and (f), we set $U$ = 5.4 eV (2$t_1$) and electron doping density $x$ = 10$\%$. }

\end{figure}
Similarly as the graphite, it is important to study the octagraphite ($n=\infty$). The DFT calculated intra-layer structure is similar as the single-layer octagraphene, with only slightly enhanced lattice size, as the interaction between the neighboring layers changes the lattice parameters slightly.

Figure~\ref{fig:bulk}(a) shows the DFT calculated band structure of octagraphite. There are always four bands near the Fermi level for a given $k_z$, which shows the 2D feature of octagraphene materials. The highest and lowest boundaries of each band are marked by $k_z$ = 0 and $k_z$ = $\pi$, respectively. The three dimensional (3D) Fermi surface has a fusiform, where the largest hole pocket is around $\Gamma$ point, see Fig.~\ref{fig:bulk}(b). It is similar to the multi-orbital Fe-based superconductor family~\cite{Hirschfeld_2011}, and shows the importance of interlayer interactions.

We here use the 3D single-orbital TB model (Eq. (\ref{eq:tb})) to construct the major band features of the octagraphite, which is given by
\begin{equation}
\label{eq:bandi}
\widetilde{H}_{\infty}=-\left[\begin{array}{cccc}{2t_{\perp}\cos k_{z}}& {t_1} & {t_2 e^{i k_{y}}+t_3} & {t_1} \\ {t_1} & {2t_{\perp}\cos k_{z}} & {t_1} & {t_2 e^{i k_{x}}+t_3} \\ {t_2 e^{i k_{y}}+t_3} & {t_1} & {2t_{\perp}\cos k_{z}} & {t_1} \\ {t_1} & {t_2 e^{-i k_{x}}+t_3} & {t_1} & {2t_{\perp}\cos k_{z}}\end{array}\right].
\end{equation}

\begin{table}[]
	\caption{\label{tab:1}%
		The lattice constant $a_0$, fitting parameters $t_1$, $t_2$, $t_3$, $t_{\perp}$ and $\lambda$ of single to six-layer of octagraphene and octagraphite ($\infty$).
	}
	\begin{ruledtabular}
		\begin{tabular}{ccccccc}
			\textrm{$n$}&
			\textrm{$a_0$ ($\AA$)}&
			\textrm{$t_1$ (eV)}&
			\textrm{$t_2$ (eV)}&
			\textrm{$t_3$ (eV)}&
			\textrm{$t_{\perp}$ (eV)}&
			\textrm{$\lambda$}\\
			\colrule
			1 		& 3.444 & 2.678(33) & 2.980(27) & 0.548(24) &  $---$	& 0.330\\
			2 		& 3.446 & 2.685(21) & 3.001(16) & 0.558(16) & 0.184(11) & 0.324\\
			3 		& 3.447 & 2.680(16) & 2.994(13) & 0.548(12) & 0.222(07) & 0.320\\
			4 		& 3.446 & 2.678(13) & 3.001(11) & 0.550(11) & 0.263(06) & 0.320\\
			5		& 3.447 & 2.671(12) & 2.993(10) & 0.546(09) & 0.261(05) & 0.320\\
			6 		& 3.449 & 2.677(11) & 2.999(09) & 0.548(08) & 0.247(05) & 0.313\\
			$\infty$& 3.447 & 2.686(17) & 2.986(13) & 0.574(12) & 0.259(05) & 0.319\\
		\end{tabular}
	\end{ruledtabular}
\end{table}

Since the $\epsilon_1$ and $\epsilon_4$ are away from the Fermi level, we only use the $\epsilon_2$ and $\epsilon_3$ with k$_z$ = 0, $\pi/2$ and $\pi$ in our fittings. By fitting the bands $\epsilon_2$ and $\epsilon_3$ from $\Gamma$ to M point, we get $t_1$ = 2.686 $\pm$ 0.017eV, $t_2$ = 2.986 $\pm$ 0.013 eV, $t_3$ = 0.574 $\pm$ 0.012 eV and $t_{\perp}$ = 0.259 $\pm$ 0.005 eV. $t_{\perp}$ here has little difference from octagraphene with layer number $n\geq$ 4.

We need now to consider the form of Fermi surface. See Fig.~\ref{fig:model}(c) from TB model Eq. (\ref{eq:tb}), the ($c_{1 \sigma}$, $c_{2 \sigma}$, $c_{3 \sigma}$, $c_{4 \sigma}$) in a unit cell can be transformed to ($-c_{1 \sigma}$, $c_{2 \sigma}$, $-c_{3 \sigma}$, $c_{4 \sigma}$) with a gauge transformation $\widetilde{T}$, like
	\begin{equation} \widetilde{T}H_{TB}(t_1,t_2,t_3,t_{\perp})\widetilde{T}^{-1}=H_{TB}(-t_1,t_2,t_3,t_{\perp}).
	\end{equation}
Since the gauge transformation $\widetilde{T}$ does not change the momentum coordinates, $H_{TB}(t_1,t_2,t_3,t_{\perp})$ would has exactly the same energy levels as $H_{TB}(-t_1,t_2,t_3,t_{\perp})$ at any momentum $\mathbf{k}$.
	
It is easily seen that when $t_3=0$ in Eq. (\ref{eq:bandi}),  $\widetilde{H}_{\infty}(\mathbf{k})$ and $\widetilde{H}_{\infty}(\mathbf{k}+(\pi,\pi,\pi))$ satisfy the following
equations,
\begin{equation}
\widetilde{H}_{\infty}(\mathbf{k},t_1,t_2,t_{\perp})=-\widetilde{H}_{\infty}(\mathbf{k}+(\pi,\pi,\pi),-t_1,t_2,t_{\perp}).
\end{equation}
Given that the eigenvalues of $\widetilde{H}_{\infty}(\mathbf{k})$ and $\widetilde{H}_{\infty}(\mathbf{k}+(\pi,\pi,\pi))$ have the same absolute value with a different sign. Consider, for simplicity, all energy levels in a half Brillouin zone must have opposite values as the other half. Therefore, the Fermi energy level is located at $E_f=0$ with half filling exactly. If eigenvalue $E_\mathbf{k}=0$ happens at a nonspecific $\mathbf{k}$, $E_\mathbf{k}$ at Fermi energy level, it is easily seen that $E_\mathbf{k+(\pi,\pi,\pi)}=0$. We finally prove the perfect Fermi surface nesting vector $\mathbf{Q}_\infty=(\pi,\pi,\pi)$ for $t_3=0$ in Eq. (\ref{eq:bandi}). When $t_3>0$, the actual Fermi surface nesting vector is deviated from $\mathbf{Q}_\infty=(\pi,\pi,\pi)$ with a limited scale.

Figure~\ref{fig:bulk}(c) shows the eigen-susceptibilities $\chi(\mathbf{q})$ for $\mathbf{q}_z$ = 0, $\pi/2$, $\pi$. $\chi(\mathbf{q})$ peaks at $\mathbf{Q}_\infty$ = $(\pi,\pi,\pi)$, and the related eigenvector of susceptibilities $\epsilon(\mathbf{Q}_\infty)$ = $(1/2,-1/2,1/2,-1/2)$ means that the N\'{e}el pattern is obtained both within the layer and between the layers with half filling, shown in Fig.~\ref{fig:bulk}(d).  The reason for that $\chi(\mathbf{q})$ peaks at $\mathbf{Q}_\infty$ = $(\pi,\pi,\pi)$ lies in that the FS- nesting vector is at $\mathbf{Q}_\infty=(\pi,\pi,\pi)$. As shown in Fig.~\ref{fig:bulk}(b), due to the inter-layer coupling, the hole pocket centering at the $\Gamma$-point is no longer nested with the electron pocket centering at the M ($\pi,\pi,0$) point with the same $k_z$, and instead it's best nested with the electron pocket centering at the ($\pi,\pi,\pi$)-point. Therefore, the FS-nesting vector is $\mathbf{Q}_\infty=(\pi,\pi,\pi)$. Note that such an inter-layer magnetic structure is new for the octagraphite and is absent for the single-layer octagraphene. What's more, the FS-nesting in this case is not perfect, which leads to a small but finite $U_c$ with half filling, see Fig.\ref{fig:bulk}(e). It means considerable superconductivity can occur even in half filling.

Finally, we get $\lambda$ = 0.319 and $T_c$ $\sim$ 170 K for the octagraphite. Practically, the $U$ of real carbon-based materials are larger than our given value $U=5.4$ eV~\cite{Schuler_2013}, this may give a chance to get a higher $T_c$ in real materials. However, the RPA given $T_c$ level is usually overestimated because of its weak-coupling perturbation, with its limitation of adopting a strong $U$~\cite{Liu_2013}. As shown in Fig. \ref{fig:bulk}(e), The RPA limited $U_c$ is above 6.0 eV when electron doping density $x$ $>$ 10$\%$.  In Fig. \ref{fig:bulk}(f), the dependence of $x$ for $\lambda$ shows that the RPA results are reliable when $U/U_c$ is far less than 1. Thus we set $U$ = 5.4 eV,  $x=$10$\%$ to approach the relatively reasonable $T_c$ in the field of our RPA limit.

We notice that $\lambda$ of octagraphite shows a small decrease from single-layer octagraphene. Note that $t_3$ here is larger than that of single-layer octagraphenes, and is negative to form the well nesting Fermi surface. The Fermi nesting is deviated by the interlayer interaction, leading to the a small decrease of $T_c$. Calculated $s^{\pm}$-wave pairing is stronger than the other three pairing symmetry channels ($p$, $d_{xy}$, $d_{x^2-y^2}$), so the superconductivity of octagraphite is also similar to multi-orbital Fe-based superconductors.
Besides, $\lambda$ of octagraphite converges to a constant value as the layer number $n\ge 3$, which means that $T_c$ changes little with $n$. This reflects the 2D nature of octagraphite.

Interestingly in Figs.~\ref{fig:single}(a), ~\ref{fig:multi}(c) and ~\ref{fig:bulk}(a), except for the four energy bands described by TB model, other bands are almost the same and independent of the layer number $n$ from the DFT results. They are represented by the local properties of orbits. Note that these bands are far away from Fermi level, so they have little influence on the superconductivity.

\section{\label{sec:con} Conclusions}

Here we study the electronic structure, magnetism and superconductivity of single-layer octagraphene, multi-layer octagraphene, and octagraphite. The DFT calculations suggest that the multi-layer octagraphene has a simple A-A stacking and the cohesive energy differences are smaller than graphene. This indicates a good slip property and a promising mechanical applications. A TB model is built to capture the main features for each layer number $n$. The hopping parameters are obtained with high accuracy. We find the hopping parameters change little with the layer number $n$. The van der Waals interaction induces $t_{\perp}\approx0.25$ eV, smaller than multi-layer graphenes. All these support that the multi-layer octagraphene and octagraphite are more 2D-like. We find the sandwich structure with the multiple energy bands overlapping frequently in the multi-layer octagraphene. This band structure has not been reported before, which may bring more interesting topological phenomena.

At the Fermi level, the band structures of octagraphenes contain hole pockets around the $\Gamma$ point and electron pockets around the $M$ point. The two pockets connected by the nesting vector $\mathbf{Q_1}=(\pi,\pi)$ form the well Fermi-surface nesting for the single-layer octagraphene. For the multi-layer octagraphene the nesting vector is blured from $\mathbf{Q}=(\pi,\pi)$, makes $T_c$ lower than the single-layer octagraphene. For octagraphite, Fermi-surface nesting is switched to 3D form with nesting vector $\mathbf{Q}_\infty=(\pi,\pi,\pi)$, also yields a high $T_c$.

By applying a RPA method with half filling, a 3D antiferromagnetic N\'{e}el magnetism is obtained both within the layer and between the layers. Thus the spin fluctuation is dominant for the SC pairing with doping. We calculate the $T_c$ of single-layer octagraphene, multi-layer octagraphene, and octagraphite, and find that the interlayer interaction would not affect the superconducting state much. With increasing the $n$, $T_c$ converges to $\sim170$ K, which is still high. The difference between the three-layer octagraphene and octagraphite is so tiny that we suggest the high-temperature superconducting $s^{\pm}$ pairing mechanism of this material is mainly a 2D mechanism.

Moreover, we find that the in-plane strain or stress would not change the energy bands obviously near the Fermi surface for the single-layer octagraphene. As an actual single-layer octagraphene may exist on a substrate, the lattice difference with the substrate would lead to some deformations. Therefore, this stability of  Fermi nesting may bring great preparation advantages. We note that the synthesis of multi-layer octagraphene is now in progress. Novel synthesis routes of multi-layer octagraphene have been reported recently~\cite{Gu_2019}. One-dimensional carbon nanoribbons with four and eight-membered rings have been synthesized experimentally~\cite{Liu_2017}. It holds great hope to realize this promising high $T_c$ material in the future.

\section{ACKNOWLEDGMENTS}
We thank Yao-Tai Kang for the RPA C++ program references, Zhihai Liu and Luyang Wang for helpful discussions. Jun Li, Shangjian Jin and Dao-Xin Yao are supported by NKRDPCGrants No. 2017YFA0206203, No. 2018YFA0306001, No. NSFC-11974432, No.  GBABRF-2019A1515011337, Leading Talent Program of Guangdong Special Projects and the start-up funding of SYSU No. 20LGPY161, Fan Yang is supported by NSFC under the Grants No. 11674025.

\nocite{*}
\bibliography{20200820}% Produces the bibliography via BibTeX.
\end{document}